\documentclass[aps,prl,showpacs,twocolumn]{revtex4}
\usepackage{graphics}
\usepackage[dvips]{color}
\usepackage{graphicx}
\usepackage{amssymb}
\usepackage{rotating}
\usepackage{epsfig}

\input epsf.sty

\newcommand{\etal}{{\em et al.}}
\newcommand{\ud}{\mbox{d}}
\newcommand{\kBT}{\mbox{$k_{_B}T$}}

\newcommand{\sbare}{\mbox{$\sigma_{\mbox{\tiny 0}}$}}
\newcommand{\sframe}{\mbox{$\sigma_{\mbox{\tiny f}}$}}
\newcommand{\sint}{\mbox{$\sigma_{\mbox{\tiny int}}$}}
\newcommand{\sfluc}{\mbox{$\sigma_{\mbox{\tiny fluc}}$}}

\begin{document}

\title{Reply to Comment on: ''Are stress-free membranes really 'tensionless'?''}
\author{Friederike Schmid}
\affiliation{Institute of Physics, JGU Mainz, D-55099 Mainz, Germany}

\begin{abstract} 
\vspace*{1cm}
\end{abstract}


\maketitle

Fournier and Barbetta state the central message in their paper as follows (see
conclusion of Ref.~\onlinecite{fournier}): ''{\em We showed that $\tau$ differs
from the tensionlike coefficient $r$ of the fluctuation spectrum and we
unveiled the correct way to derive $\tau$ from the free energy.}'' Here $\tau$
is the frame tension, called $\sframe$ in my work, and $r$ is the fluctuation
tension, called $\sfluc$ in my work. In Ref.\ \onlinecite{fs}, I argue that the
arguments of Fournier and Barbetta -- as well as those of other previous
authors who came to the same conclusion\cite{imparato,stecki} -- are inherently
inconsistent, due to the fact that they are based on a theory which has been
linearized with respect to a small parameter $(A-A_p)/A_p$, yet predict an
effect which is nonlinear in this parameter.  (Here $A$ is the membrane area
and $A_p$ the projected membrane area.) Fournier now claims that the arguments
in Ref.\ \onlinecite{fournier} is really based on an expansion in $\kBT$, which
would be the usual expansion in a diagrammatic field-theoretic treatment of the
problem. 

Let us therefore consider the expansion in $\kBT$. The starting point is the
Helfrich Hamiltonian ${\cal H}_{\mbox{\tiny Helfrich}}$ in Monge gauge, {\em
i.e.}, a gauge where the coordinates $(x,y)$ are defined by projection onto 
the plane of the frame and the membrane is parametrized by the normal distance
$h(x,y)$ onto this plane. In the field-theoretic treatment, 
${\cal H}_{\mbox{\tiny Helfrich}}$ is expanded about a planar reference state, 
yielding
\cite{kleinert}
\begin {equation}
{\cal H}= {\cal H}_0 + {\cal H}'
\end{equation}
with
\begin{eqnarray}
{\cal H}_0 &=& \sbare A_p + \frac{1}{2} \int_{A_p}
 \ud A_p \{ \sbare (\nabla h)^2 + \kappa (\Delta h)^2 \}, \\
{\cal H}' &=&
- \int_{A_p} \ud A_p \Big[
\frac{\sbare}{8} (\nabla h)^4
+ \frac{\kappa}{2} \big(\frac{1}{2} (\nabla h)^2 (\Delta h)^2
\nonumber \\ &&
\quad + \: 2 (\Delta h) (\partial_\alpha h)(\partial_\beta h)
(\partial_\alpha \partial_\beta h) +  \cdots
\Big].
\end{eqnarray}
Here $\sbare$ is the bare tension, which coincides with the internal tension
$\sint$ in a system with fixed number of lipids, and $\kappa$ is the bending
rigidity. The Hamiltonian ${\cal H}_0$ is Gaussian and ${\cal H}'$ subsumes the
nonlinear terms.  The $\kBT$ expansion is based on the full nonlinear
Hamiltonian ${\cal H}$, where ${\cal H}'$ is treated as a perturbation about
${\cal H}_0$ in a diagrammatic scheme.  Further corrections come in through the
nonlinearity of the measure ${\cal D}[h]$ \cite{cai}.  Within this approach,
some quantities can be evaluated up to first order $\kBT$ without having to
actually consider nonlinear corrections to ${\cal H}_0$.  The frame tension
$\sframe$ is probably such a quantity, and therefore, Eq.\ (4) in Ref.\
\onlinecite{fs} indeed gives the correct leading correction for $\sframe/\sint$
in a $\kBT$ expansion. The fluctuation tension $\sfluc$, however, is {\em not}
such a quantity \cite{cai}, and the calculation of the $\kBT$ order involves
the calculation of first-order loop diagrams. Such calculations can be very
tricky \cite{cai} and have not been attempted in Refs.\ \onlinecite{fournier,
imparato, stecki}. If Fournier and Barbetta meant to show $\sframe \ne \sfluc$
by an expansion in powers of $\kBT$, then their calculation not just
inconsistent, it is incomplete. In that case, they should have finished
the one-loop calculation for $\sfluc$ before making any claims.

We conclude that the arguments of Refs.\ \onlinecite{fournier,imparato,stecki}
clearly cannot be justified by an expansion in $\kBT$ or the corresponding
dimensionless quantity\cite{note1} $\epsilon=\kBT/\kappa$. Instead, the authors
Refs.\ \onlinecite{fournier,imparato,stecki} have simply replaced ${\cal H} =
{\cal H}_0$, which implies (among other) omitting higher order terms in
$(\nabla h)^2 \ll 1$ and {\em setting} $\sfluc=\sint$.  This is the
approximation examined in Ref.\ \onlinecite{fs}.  From the relation $\ud A =
\sqrt{(1-(\nabla h)^2} \: \ud A_p$, one gets locally
\begin{displaymath} 
(\nabla h)^2 = \frac{(\ud A)^2 - (\ud A_p)^2}{(\ud A_p)^2} = \eta (2 + \eta) 
\end{displaymath} 
with $\eta(x,y) = \ud(A-A_p)/\ud A_p$, hence $(\nabla h)^2 \ll 1$ implies $\eta
\ll 1$. The global average of $\eta$ is $\bar{\eta} = (A-A_p)/A_p$,  which is
thus a small parameter in {\em this} approximation: It neglects terms that are
not linear in $\bar{\eta}$. The ''expansion'' is not systematic, because other
terms (higher orders of higher derivatives of $h$) are neglected as well, but
this is not important for our argument.

It is important ot note that the parameters $\bar{\eta}= (A-A_p)/A_p$ and
$\epsilon = \kBT/\kappa$ can be varied independently. This is physically
feasible, since $A_p$ can be controlled either directly or by tuning the frame
tension, independent of the temperature $\kBT$.  The approximations $\bar{\eta}
\ll 1$ and $\epsilon \ll 1$ are thus not equivalent.  On the one hand,
capillary wave Hamiltonians \cite{cw} that ignore bending terms -- which
corresponds to setting $\kappa \to 0$ or $\epsilon \to \infty$ -- have been
extremely successful in describing the properties of liquid/liquid interfaces
at large wavelengths. On the other hand, membranes with fixed number of lipids
and approximately fixed area per lipid can be studied at fixed projected area.
This is actually a common setting in simulations. 

By appropriate Legendre transforms, one can calculate the free energy of the
Gaussian model ${\cal H}_0$ in such a $(N,A,A_p)$ ensemble:
\begin{eqnarray}
\frac{F(N,A,A_p)}{\kBT (N-1)} &=& - \frac{1}{2} \Big[
\ln\big(1-\exp(-\frac{8 \pi \kappa}{\kBT} \: \frac{(A-A_p)}{A_p})\big)
\nonumber \\ &&
+ \ln\big(\frac{A_p \kBT}{8 \pi \kappa \lambda^2 (N-1) }\big) + 2 \Big].
\end{eqnarray}
The frame tension and the fluctuation tension tension can be calculated {\em
via} $\sframe = \partial F/\partial A_p$ and $\sfluc=\sint = -\partial
F/\partial A$, and the results are of course the same as those presented in
Refs.\ \onlinecite{fs, fournier, imparato,stecki} for other ensembles.
Nevertheless, the results from the Gaussian model clearly cannot be trusted at
order $((A-A_p)/A_p)^2$.  Nonlinear effects will become important even at small
temperatures if $(A-A_p)/A_p$ is large. 

I wish to stress once more that this whole controversy is not about the
relation between the frame tension and the {\em internal} tension, but about
the {\em fluctuation} tension. One should give Farago and Pincus credit for
having been the first to derive the relation (4) in Ref.\ \onlinecite{fs}
between $\sframe$ and $\sint$ for compressible membranes with fixed number of
lipids \cite{farago1}. As Fournier correctly points out, this result gives most
likely the correct leading order in a diagrammatic expansion in powers of
$\kBT/\kappa$. However, $\sint$ is a rather uninteresting quantity, since it
can neither be controlled nor measured. Farago and Pincus recognized that their
result does not carry over to the fluctuation tension, and gave a very general
argument why the fluctuation tension should equal the frame tension
\cite{farago2,farago3}, which solely relies on the requirement of ''rotational
invariance'', {\em i.e.}, gauge invariance.  Their reasoning is similar to  a
classic argument by Cai \etal \cite{cai}, who showed $\sfluc=\sframe$ for
incompressible membranes with variable number of lipids.  The conclusion that
gauge invariance leads to $\sfluc=\sframe$ has very recently been corroborated
by numerical simulations \cite{farago3}. 

Notwithstanding, the highly accurate simulations of Ref.\ \onlinecite{fs} 
suggest that $\sfluc$ should be slightly renormalized, $\sfluc=\sframe (A_p/A)$. 
This result is in line with model-free thermodynamic considerations 
on the relation between different tension parameters in vesicles \cite{diamant}. 
Whether and how it can be reconciled with the general arguments for
$\sfluc=\sframe$ quoted above still remains to be elucidated.


\begin{thebibliography}{66}
\bibitem{note1} The parameter $\epsilon=\kBT/\kappa$ is the 
only dimensionless quantity which can be constructed from 
$\kBT, \sigma_0$, and $\kappa$ and which is linear in $\kBT$.
\end{thebibliography}
\end{document}